\documentclass[pra,twocolumn,showpacs,superscriptaddress,amssymb,floatfix]{revtex4}
\usepackage[dvips]{graphicx}
\usepackage{bm}
\newcommand{\Journal}[4]{#1 {\bf #2}, #3 (#4)}
\newcommand{\PR}{Phys. Rev.}
\newcommand{\PRL}{Phys. Rev. Lett.}
\newcommand{\PRA}{Phys. Rev. A}
\newcommand{\JMP}{J. Math. Phys.}

\newcommand{\PLA}{Phys. Lett. A}

\begin{document}
\title {Theory of spinor Fermi and Bose gases in tight atom waveguides} 
\author{M. D. Girardeau}
\email{girardeau@optics.arizona.edu}
\affiliation{Optical Sciences Center, University of Arizona, Tucson, AZ 85721}
\author{M. Olshanii}
\email{olshanii@phys4adm.usc.edu}
\affiliation{Department of Physics and Astronomy, University of Southern 
California, Los Angeles, CA 90089-0484}
\date{\today}
\begin{abstract}
Divergence-free pseudopotentials for spatially even and odd-wave
interactions in spinor Fermi gases in tight atom waveguides are derived.
The Fermi-Bose mapping method is used to relate the effectively
one-dimensional fermionic many-body problem to that of a spinor Bose 
gas. Depending on the relative magnitudes of the even and odd-wave
interactions, the $N$-atom ground state may have total spin $S=0$, $S=N/2$,
and possibly also intermediate values, the case $S=N/2$ applying near a p-wave
Feshbach resonance, where the $N$-fermion ground state is space-antisymmetric
and spin-symmetric. In this case the fermionic ground state maps to 
the spinless bosonic Lieb-Liniger gas. An external magnetic field with a 
longitudinal gradient causes a Stern-Gerlach spatial separation of the 
corresponding trapped Fermi gas with respect to various values of $S_z$.
\end{abstract}
\maketitle
\section{Introduction}
When an ultracold atomic vapor is placed into an atom waveguide with
sufficiently tight transverse confinement, its two-body scattering
properties are strongly modified. This occurs in a regime of low temperatures
and densities where transverse oscillator modes are frozen and the dynamics 
is described by a one-dimensional (1D) Hamiltonian with zero-range 
interactions \cite{Ols98,PetShlWal00}, a regime which has already been reached 
experimentally \cite{Gre02,MorStoKohEss03,Tol03}. In addition, a regime 
with chemical potential $\mu$ less than transverse level 
spacing $\hbar\omega_{\perp}$ but $k_{B}T>\hbar\omega_{\perp}$ 
has been achieved \cite{Bon01,GorVogLea01}.
We assume herein that both $\mu<\hbar\omega_{\perp}$ and 
$k_{B}T<\hbar\omega_{\perp}$ as in \cite{Gre02,MorStoKohEss03,Tol03}.
Nevertheless, \emph{virtually} excited transverse 
modes renormalize the effective 1D coupling constant via a confinement-induced
resonance, as first shown for bosons \cite{Ols98,BerMooOls03} and recently 
for spin-polarized fermionic vapors by Granger and Blume \cite{GraBlu03}.

The dynamics of an optically trapped Fermi gas is richer than that of
a magnetically trapped one, since the spin is not polarized. A Fermi-Bose
mapping first used to solve the 1D hard-sphere Bose gas \cite{Gir60,Gir65} 
was recently shown \cite{CheShi98} to provide an exact
duality between effective zero-range 1D fermionic and bosonic interactions and 
applied to spin-polarized Fermi gases \cite{GraBlu03,GirOls03}. 
It will be shown here that this mapping and Fermi-Bose duality  
also hold for spinor Fermi gases. This mapping will be exploited to reduce 
the degenerate spatially antisymmetric fermionic ground states to that of a 
spinless Bose gas, which, in the 1D, zero-range
interaction regime, is the Lieb-Liniger model
which is exactly soluble in the absence of longitudinal trapping 
\cite{LieLin63,KorBogIze93} and well approximated by a local equilibrium
approach in the trapped case \cite{DunLorOls01}. It will be shown 
that in the presence of longitudinal trapping, this fermionic ground state  
can be
Stern-Gerlach spatially decomposed by a longitudinal magnetic field 
gradient into components with various values of total longitudinal spin.   
\section{Two-body problem for fermions in a tight waveguide} 
Consider first the
three-dimensional two-body scattering problem for spin-$\frac{1}{2}$ fermionic
atoms. There are both s-wave scattering states, which are space
symmetric and spin antisymmetric with spin eigenfunctions of singlet
form $\frac{1}{\sqrt{2}}(\uparrow\downarrow-\downarrow\uparrow)$, 
as well as p-wave scattering states which are space antisymmetric 
and spin symmetric with spin eigenfunctions
of triplet form $\uparrow\uparrow$ or $\downarrow\downarrow$ or
$\frac{1}{\sqrt{2}}(\uparrow\downarrow+\downarrow\uparrow)$. s-wave scattering
cannot occur in a spin-polarized Fermi gas, but it is usually dominant in
a spinor Fermi gas since p-wave spatial antisymmetry suppresses short-range
interactions. However, both s-wave and p-wave interactions can be greatly 
enhanced by Feshbach resonances \cite{Rob01,RegTicBohJin03}.
Assume until further notice that the Hamiltonian does not depend on spin.
Then the spin dependence of wave functions need not be
indicated explicitly and they can be written as the sum of  spatially even
and odd parts $\psi_e$ and $\psi_o$. When such an atomic vapor is
confined in an atom waveguide with tight transverse trapping, the dynamics 
becomes effectively 1D \cite{Ols98}. 
The effective 1D interactions are determined by 1D
scattering lengths $a_{1D}^e$ for spatially even waves 
$\psi_{e}(z)=\psi_{e}(-z)$ related to 3D s-wave scattering and spatially
odd waves $\psi_{o}(z)=-\psi_{o}(-z)$ related to 3D p-wave scattering;
here $z$ is the relative coordinate $z_{1}-z_{2}$ for 1D scattering. These
determine the $k\to 0$ behavior 
just outside the range $z_0$ of the interaction:
\begin{eqnarray}\label{contact}
\psi_{e}^{'}(z_0)
&=&-\psi_{e}^{'}(-z_0)=-(a_{1D}^{e}-z_{0})^{-1}\psi_{e}(\pm z_0)\nonumber\\
\psi_{o}(z_0)&=&-\psi_{o}(-z_0)
=-(a_{1D}^{o}-z_{0})\psi_{o}^{'}(\pm z_0)\quad .
\end{eqnarray}
$a_{1D}^e$ is a known \cite{Ols98} function of the 3D s-wave scattering
length $a_s$, and $a_{1D}^o$ is a known \cite{GraBlu03,Note1} function of the 
3D p-wave scattering volume 
$V_{p}=a_{p}^{3}=-\lim_{k\to 0}\tan\delta_{p}(k)/k^3$ \cite{SunEsrGre03}:
\begin{eqnarray}\label{renorm}
(a_{1D}^e)^{-1}&=&\frac{-2a_s}{a_{\perp}^2}[1-(a_{s}/a_{\perp})
|\zeta(1/2)|]^{-1}\nonumber\\ 
a_{1D}^{o}&=&\frac{6V_{p}}{a_{\perp}^2}[1+12(V_{p}/a_{\perp}^3)
|\zeta(-1/2,1)|]^{-1} 
\end{eqnarray}
where $a_{\perp}=\sqrt{\hbar/\mu\omega_{\perp}}$ is the transverse oscillator
length for the relative motion, $\mu$ is the effective mass, 
$\zeta(1/2)=-1.460\cdots$ is a Riemann zeta function, and 
$\zeta(-1/2,1)=-\zeta(3/2)/4\pi=-0.2079\cdots$ is a Hurwitz zeta function
\cite{WhiWat52}. 

In the zero-range limit $z_0$ approaches $0+$ and
$-z_0$ approaches $0-$, and Eqs. (\ref{contact}) reduce to ``contact
conditions'' which relate a discontinuity in $\psi_{e}^{'}$ at contact
$z=0$ to $\psi_{e}(0)$, and a discontinuity of $\psi_{o}$ to
$\psi_{o}^{'}(0)$. Although a discontinuity in the derivative 
is a well-known consequence of the zero-range delta function pseudopotential
and plays a crucial role in the solution of the Lieb-Liniger model
\cite{LieLin63}, discontinuities of $\psi$ itself have received 
little attention, although they have been discussed previously by
Cheon and Shigehara \cite{CheShi98} and are implicit in the recent work
of Granger and Blume \cite{GraBlu03}. For an odd wave $\psi_o$ the 
discontinuity $2\psi(0+)$ is a trivial consequence of antisymmetry 
together with the fact that a nonzero odd-wave
scattering length cannot be obtained in the limit $z_{0}\to 0$ unless 
$\psi_{o}(0\pm)\ne 0.$ These discontinuities are rounded off when $z_{0}>0$,
since the interior wave function interpolates smoothly between the values
at $z=-z_0$ and $z=z_0$. 
A general 1D two-body wave function $\psi(z)$ is the sum
of even and odd parts: $\psi(z)=\psi_{e}(z)+\psi_{o}(z)$, and the zero-range
limit of Eqs. (\ref{contact}) can be combined into 
\begin{eqnarray}\label{General-contact}
\psi'(0+)-\psi'(0-)= -(a_{1D}^{e})^{-1}[\psi(0+) + \psi(0-)]\nonumber
\\
\psi(0+)-\psi(0-)= -a_{1D}^{o}[\psi'(0+) + \psi'(0-)]\quad .
\end{eqnarray}
\section{Even and odd-wave pseudopotentials} 
Take the Hamiltonian to be 
\begin{equation}\label{H_1D}
\hat{H}_{1D}=-(\hbar^{2}/2\mu)\partial_{z}^{2}+v_{1D}^{e}+v_{1D}^{o}
\end{equation}
where $v_{1D}^{e}$ and $v_{1D}^{o}$ are even- and odd-wave pseudopotentials
to be determined. $\partial_{z}^{2}$ is nonsingular for $z\ne 0$,  
but at the origin there are singular contributions. The first 
derivative is
$\partial_{z}\psi(z)=\psi^{'}(z\ne 0)+[\psi(0+)-\psi(0-)]\delta(z)$.
The second derivative then has two contributions in addition to 
$\psi^{''}(z\ne 0)$, one because in general $\psi^{'}(0+)\ne\psi^{'}(0-)$
and the other from the derivative of the delta function:
\begin{eqnarray}\label{KE}
\partial_{z}^{2}\psi(z)&=&\psi^{''}(z\ne 0)
+[\psi^{'}(0+)-\psi^{'}(0-)]\delta(z)\nonumber\\
&+&[\psi(0+)-\psi(0-)]\delta^{'}(z)\ .
\end{eqnarray}
The $\delta(z)$ term is standard in the theory of zero-range even-wave 
interactions, the derivative discontinuity in $\psi(z)$ being chosen to
cancel a zero-range even-wave interaction proportional to $\delta(z)$,
but the $\delta^{'}(z)$ term is new. To cancel it an
odd-wave pseudopotential proportional to $\delta^{'}(z)$ suggests itself.
However, the discontinuity in $\psi$ leads to unwanted products of
delta functions unless regularizing operators are included.
Define two linear operators $\hat{\delta}_{\pm}$ and $\hat{\partial}_{\pm}$ by
\begin{eqnarray}\label{exterior}
\hat{\delta}_{\pm}\psi(z)&=&(1/2)[\psi(0+)+\psi(0-)]\delta(z)\nonumber\\
\hat{\partial}_{\pm}\psi(z)&=&(1/2)[\psi^{'}(0+)+\psi^{'}(0-)]
\end{eqnarray}
where $\delta(z)$ is the usual Dirac delta function. The divergence-free
even and odd-wave pseudopotential operators are then 
\begin{equation}\label{pseudopotentials}
v_{1D}^{e}=g_{1D}^{e}\hat{\delta}_{\pm}\quad ,\quad 
v_{1D}^{o}=g_{1D}^{o}\delta^{'}(z)\hat{\partial}_{\pm}\ .
\end{equation}
They satisfy convenient projection
properties $v_{1D}^{e}\psi_{o}=v_{1D}^{o}\psi_{e}=0$ on the even and odd
parts of $\psi$, and their matrix elements are 
$\langle\chi|v_{1D}^{e}|\psi\rangle=\frac{1}{2}\chi^{*}(0)[\psi(0+)+\psi(0-)]$
and $\langle\chi|v_{1D}^{o}|\psi\rangle
=-\frac{1}{2}[\chi^{'}(0)]^{*}[\psi^{'}(0+)+\psi^{'}(0-)]$. They are 
unambiguous and connect only even to even and odd to odd wave functions
if we stipulate that $\chi(0)=0$ [the average of $\chi(0+)$ and $\chi(0-)$]
if $\chi$ is odd and $\chi^{'}(0)=0$ [the average of
$\chi^{'}(0+)$ and $\chi^{'}(0-)$] if $\chi$ is even. In fact, the wave
function and its derivative at $z=0$ refer to the \emph{internal} wave
function as modified by the potential, whereas $z=0+$ and
$z=0-$ refer to the wave function \emph{just outside} the range of the
potential, and the above values at $z=0$ follow from the way the internal
wave function interpolates between the contact conditions on the
\emph{exterior} wave function (see below).
Terms in $\delta(z)$ and $\delta^{'}(z)$ cancel from
$\hat{H}_{1D}$ if
\begin{eqnarray}
g_{1D}^{e}[\psi(0+)+\psi(0-)]
&=&(\hbar^{2}/\mu)[\psi^{'}(0+)-\psi^{'}(0-)]\nonumber\\
g_{1D}^{o}[\psi^{'}(0+)+\psi^{'}(0-)]
&=&(\hbar^{2}/\mu)[\psi(0+)-\psi(0-)
\end{eqnarray}
and these are equivalent to the contact conditions (\ref{General-contact})
if $g_{1D}^{e} = -\hbar^{2}/\mu a_{1D}^e$ and 
$g_{1D}^{o} = -\hbar^{2}a_{1D}^{o}/\mu$. 

The physical significance is clarified by starting
from a nonsingular square well. Take the potential 
$v(z)$ to be $-V_0$ when 
$-z_{0}<z<z_0$ and zero when $|z|>z_0$. (The odd-wave
interaction $v_{1D}^o$ in $\hat{H}_{1D}$ is \emph{negative}
definite in the regime of interest, where $g_{1D}^{o}>0$.) The antisymmetric
solution $\psi_{o}$ of the zero-energy scattering equation
$[(-\hbar^{2}/2\mu)\partial_{z}^{2}+v(z)]\psi_{o}(z)=0$ inside the well 
is $\sin(\kappa z)$ with $\kappa=\sqrt{2\mu V_{0}/\hbar^2}$. The odd-wave
scattering length $a_{1D}^{o}$ is defined by the second Eq. (\ref{contact}), 
which is satisfied in the limit $z_{0}\to 0+$ if $V_0$ scales with $z_0$ as 
$\kappa=(\pi/2z_{0})[1+(2/\pi)^2 (z_{0}/a_{1D}^{o})]$. In that limit the
boundary conditions reduce to the second
Eq. (\ref{General-contact}). Inside the well the kinetic and potential
energy terms are $-(\hbar^{2}/2\mu)\partial_{z}^{2}\psi_{o}(z)
=-(\hbar^{2}\kappa^{2}/2\mu)\sin(\kappa z)$ and 
$v(z)\psi_{o}(z)=-V_{0}\sin(\kappa z)$. For $|z|<z_0$,
$\cos(\kappa z)$ is proportional to a representation of $\delta(z)$
as $z_{0}\to 0$, since 
$\int_{-z_{0}}^{z_{0}}\cos(\kappa z)f(z)dz
\to f(0)\int_{-z_{0}}^{z_{0}}\cos(\kappa z)dz
=f(0)2\kappa^{-1}\sin(\kappa z_{0})\to 2z_{0}f(0)$.  
Then its derivative $-(\kappa/2z_{0})\sin(\kappa z)$
is a representation of $\delta^{'}(z)$. Noting that $\kappa z_{0}\to\pi/2$
as $z_{0}\to 0$ we have 
$-(\hbar^{2}/2\mu)\partial_{z}^{2}\psi_{o}(z)
=-(\hbar^{2}\kappa^{2}/2\mu)\sin(\kappa z)\to 
(\pi\hbar^{2}/2\mu)\delta^{'}(z)$ which agrees with the kinetic energy term
$-(\hbar^{2}/2\mu)[\psi_{o}(0+)-\psi_{o}(0-)]\delta^{'}(z)$ from
Eq. (\ref{KE}) since $\psi_{o}(0+)$ and 
$\psi_{o}(0-)$ are to be interpreted as $\psi_{o}(z_{0})$ and 
$\psi_{o}(-z_{0})$ as $z_{0}\to 0+$. Next consider the
potential energy term inside the well as $z_{0}\to 0+$:
$-V_{0}\sin(\kappa z)\to -V_{0}(-2z_{0}/\kappa)\delta^{'}(z)
\to(\pi\hbar^{2}/2\mu)\delta^{'}(z)$. Comparing this with 
$v_{1D}^{o}\psi_{o}(z)$ from Eq. (\ref{pseudopotentials}), using the
expression for $g_{1D}^{o}$, noting that $\psi^{'}(0\pm)$ in 
Eq. (\ref{exterior}) are to be interpreted as $\psi^{'}(\pm z_{0})$,
one finds that the two expressions for the potential energy term agree
in the limit $z_{0}\to 0+$. 
\section{Fermi-Bose mapping} 
The two-body states $\psi(z)$ considered
so far are fermionic, i.e., the spatially even part $\psi_{e}(z)$ contains
an implicit spin-odd singlet spin factor, and the spatially odd part
$\psi_{o}(z)$ contains implicit spin-even triplet spin factors. To emphasize
the combined space-spin fermionic antisymmetry, these will now be denoted
by $\psi_{F}(z)=\psi_{F}^{e}(z)+\psi_{F}^{o}(z)$. States of combined space-spin
bosonic symmetry can be defined by the mapping 
$\psi_{B}(z)=\text{sgn}(z)\psi_{F}(z)$ where $\text{sgn}(z)$ is $+1$ if 
$z>0$ and $-1$ if $z<0$. This maps the spatially even fermionic function
$\psi_{F}^{e}$ to a spatially odd bosonic function $\psi_{B}^{o}$ and
the spatially odd fermionic function $\psi_{F}^{o}$ to a spatially even
bosonic function $\psi_{B}^{e}$ while leaving the spin dependence unchanged,
and the corresponding scattering lengths are also unchanged: 
$a_{1D,B}^{o}=a_{1D,F}^{e}$ and $a_{1D,B}^{e}=a_{1D,F}^{o}$. Then 
the even-wave contact conditions for $a_{1D,B}^{e}$ follow from the 
odd-wave contact conditions for $a_{1D,F}^{o}$ and the odd-wave contact 
conditions for $a_{1D,B}^{o}$ follow from the even-wave contact conditions for
$a_{1D,F}^{e}$. Since the kinetic energy contributions from $z\ne 0$ also 
agree, one has a mapping from the fermionic to bosonic problem which 
preserves energy eigenvalues and dynamics. The bosonic Hamiltonian is of
the same  form as the fermionic one (\ref{H_1D}) but with mapped coupling
constants $g_{1D,B}^{e}=\hbar^{4}/\mu^{2}g_{1D,F}^{o}$ and 
$g_{1D,B}^{o}=\hbar^{4}/\mu^{2}g_{1D,F}^{e}$, the first of which agrees
with the low-energy limit of Eq. (25) of \cite{GraBlu03,Note1}. 
In the limit $g_{1D,B}^{e}=+\infty$ arising when $V_{p}\to 0-$, this is
the $N=2$ case of the original mapping \cite{Gir60,Gir65} from hard sphere 
bosons to an ideal Fermi gas, now generalized to arbitrary
coupling constants and spin dependence. This generalizes to
arbitrary $N$: Fermionic solutions 
$\psi_{F}(z_{1},\sigma_{1};\cdots;z_{N},\sigma_{N})$
are mapped to bosonic solutions 
$\psi_{B}(z_{1},\sigma_{1};\cdots;z_{N},\sigma_{N})$ via
$\psi_{B}
=A(z_{1},\cdots,z_{N})\psi_{F}(z_{1},\sigma_{1};\cdots;z_{N},\sigma_{N})$
where $A=\prod_{1\le j<\ell\le N}\text{sgn}(z_{j\ell})$
is the same mapping function used originally \cite{Gir60,Gir65} and
the spin z-component arguments $\sigma_j$ take on the values
$\uparrow$ and $\downarrow$.  
The N-fermion and N-boson Hamiltonians are both of the form
$\hat{H}_{1D}=-(\hbar^2/2m)\sum_{j=1}^{N}\partial_{z_j}^{2}
+\sum_{1\le j<\ell\le N}[g_{1D}^{e}\hat{\delta}_{j\ell}
+g_{1D}^{o}\delta^{'}(z_{j\ell})\hat{\partial}_{j\ell}]$
generalizing (\ref{H_1D}) and (\ref{pseudopotentials}), where the linear
operators $\hat{\delta}_{j\ell}$ and $\hat{\partial}_{j\ell}$ are
defined on the Hilbert space of N-particle wave functions $\psi$ by
$\hat{\delta}_{j\ell}\psi=(1/2)[\psi|_{z_{j}=z_{\ell +}}
+\psi|_{z_{j}=z_{\ell -}}]\delta(z_{j}-z_{\ell})$ and
$\hat{\partial}_{j\ell}\psi=(1/2)[\partial_{z_{j}}\psi|_{z_{j}=z_{\ell +}}
-\partial_{z_{\ell}}\psi|_{z_{j}=z_{\ell -}}]$. On fermionic states $\psi_F$,
$g_{1D}^{e}$ and 
$g_{1D}^{o}$ are $g_{1D,F}^{e}$ and $g_{1D,F}^{o}$, whereas on the mapped
bosonic states $\psi_{B}=A\psi_F$ they are 
$g_{1D,B}^{e}=\hbar^{4}/\mu^{2}g_{1D,F}^{o}$ 
and $g_{1D,B}^{o}=\hbar^{4}/\mu^{2}g_{1D,F}^{e}$.
\section{N-particle ground state} 
\begin{figure}
\includegraphics[width=1.0\columnwidth,angle=0]{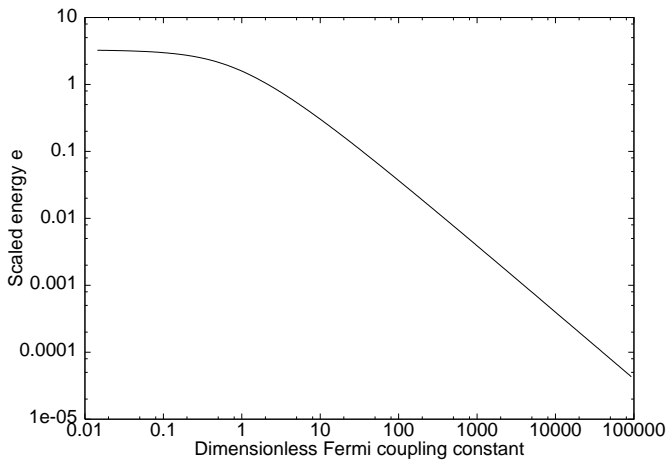}
\caption{Log-log plot of scaled ground state energy per particle 
$e=2m\epsilon/\hbar^{2}n^2$ for the spatially antisymmetric spinor Fermi gas, 
versus dimensionless fermionic coupling constant $\gamma_F$.}
\label{Fig:one}
\vspace{-0.5cm}
\end{figure}

Assume that both $g_{1D,F}^{e}\ge 0$ and 
$g_{1D,F}^{o}\ge 0$. If $g_{1D,F}^{o}$ is zero or negligible, 
then it follows from a theorem of Lieb and Mattis \cite{LieMat62} that
the fermionic ground state has total spin $S=0$ (assuming $N$ 
even), as shown in the spatially uniform case by Yang \cite{Yan67} and 
with longitudinal trapping by Astrakharchik {\it et al.} 
\cite{AstBluGioPit03}. 
If $g_{1D,F}^{o}$ is not negligible then the ground state may not have $S=0$. 
In fact, if $g_{1D,F}^{e}$ is zero or negligible then one can apply a
theorem of Eisenberg and Lieb \cite{EisLie02} to the mapped spinor boson
Hamiltonian, with the conclusion that the degenerate Bose ground state is
totally spin-polarized, has $S=N/2$, and is the product of a symmetric
spatial wave function $\psi_{B0}$ and a symmetric spin wave function. 
$\psi_{B0}$ is the ground state of the Lieb-Liniger gas 
\cite{LieLin63,KorBogIze93} which is known for all positive $g_{1D,B}^{e}$ 
(hence all positive mapped $g_{1D,F}^{o}$) in the absence of longitudinal 
trapping. The inverse mapping then yields the 
$N$-fermion ground state, which has a totally space-antisymmetric and 
spin-symmetric wave function, which is $(N+1)$-fold degenerate 
since $S_z$ ranges from $-N/2$ to $N/2$. 
Define dimensionless bosonic and fermionic coupling constants by 
$\gamma_{B}=mg_{1D,B}^{e}/n\hbar^2$ and 
$\gamma_{F}=mg_{1D,F}^{o}n/\hbar^2$ where $n$ is the longitudinal
particle number density. They satisfy $\gamma_{B}\gamma_{F}=4$. 
The energy per particle $\epsilon$ is
related to a dimensionless function $e(\gamma)$ available online
\cite{Note2} via $\epsilon=(\hbar^{2}/2m)n^{2}e(\gamma)$ where $\gamma$
is related to $\gamma_F$ herein by $\gamma=\gamma_{B}=4/\gamma_{F}$.
This is plotted as a function of $\gamma_F$ in Fig. \ref{Fig:one}.
For $g_{1D,F}^{o}\to\infty$ as occurs at a p-wave Feshbach resonance,
one has a ``fermionic TG gas'' \cite{GirOls03}  mapping to a zero-energy 
\emph{ideal Bose} gas, a fermionic analog of the 
``TG gas'' of impenetrable point bosons mapping to an \emph{ideal Fermi} 
gas \cite{Gir60,Gir65,Ols98,GorVogLea01,Gre02,LieSeiYng03}. Any $S=0$ state 
has a higher energy in this case; in fact, for $N>2$ the mapped Bose gas is 
partially space-antisymmetric, raising its energy by the exclusion principle.

In the presence of a uniform external magnetic field $h$ the directional
degeneracy is lifted and the above $N$-particle state is the (now 
nondegenerate) ground state with field quantization direction parallel to
the field, and the ground state energy is lowered by an amount 
$N\mu_{\text{B}}h/2$ where $\mu_{\text{B}}h/2$ is the magnetic moment of 
each spin-$\frac{1}{2}$ atom.

So far we have considered only the extremes of an $S=0$ ground state 
(large $g_{1D,F}^{e}$) or one with $S=N/2$ (large
$g_{1D,F}^{o}$). The determination of the state of lowest energy 
for arbitrary values of these coupling constants is as yet only partially
solved, although we have recently obtained exact results for the phase
diagram of ground-state total spin \cite{GirOls04-2}.
\section{Response to a magnetic field gradient} 
Suppose that the spinor
Fermi gas is longitudinally trapped by an optical potential
$\hat{V}_{\text{trap}}=\sum_{j}\frac{1}{2}m\omega_{\text{long}}^{2}z_{j}^2$, 
and that there is also a 
longitudinal magnetic field $h(z)=cz$ with constant gradient $c$, adding 
an interaction term 
$\hat{V}_{\text{space-spin}}=-\mu_{\text{B}}c\sum_{j}\hat{s}_{jz}z_{j}$ to 
the $N$-particle Hamiltonian $\hat{H}_{1D}^{F}$, where $\hat{s}_{jz}$ is the
spin z-component operator for the $j${\it th} particle. The space-spin 
interaction terms can be eliminated by a canonical transformation 
$\hat{U}^{-1}z_{j}\hat{U}=z_{j}-\alpha \hat{s}_{jz}$, 
$\hat{U}^{-1}\hat{p}_{j}\hat{U}=\hat{p}_{j}$, 
$\hat{U}^{-1}\hat{s}_{jz}\hat{U}=\hat{s}_{jz}$
which leave the canonical commutation relations invariant. Noting that
$\hat{U}^{-1}\hat{V}_{\text{space-spin}}\hat{U}
=\hat{V}_{\text{space-spin}}+\frac{1}{4}N\mu_{\text{B}}c\alpha$ 
and $\hat{U}^{-1}\hat{V}_{\text{trap}}\hat{U}
=\sum_{j}\frac{1}{2}m\omega_{\text{long}}^{2}(z_{j}-\alpha \hat{s}_{jz})^2$,
one finds that the space-spin coupling terms cancel from 
$\hat{U}^{-1}\hat{H}_{1D}^{B}\hat{U}$ with the
choice $\alpha=-\mu_{\text{B}}c/(m\omega_{\text{long}}^{2})$, leading to
a transformed Hamiltonian 
$\hat{U}^{-1}\hat{H}_{1D}^{B}\hat{U}=\hat{H}_{1D}^{B}(c=0)
-N(\mu_{\text{B}}c)^{2}/(8m\omega_{\text{long}}^{2})$ where 
$\hat{H}_{1D}^{B}(c=0)$ does not include 
$\hat{V}_{\text{space-spin}}$. The ground state 
$\phi_{B0}(z_{1},s_{1};\cdots,z_{N},s_{N})$ of 
$\hat{U}^{-1}\hat{H}_{1D}^{B}\hat{U}$
is the same as that of $\hat{H}_{1D}^{B}(c=0)$. The corresponding 
single-particle
density $n_{0}(z)$ is centered on $z=0$. It is not known analytically
in the presence of longitudinal trapping, but accurate numerical results
have been calculated by a local
density method \cite{DunLorOls01}. In the presence of 
$\hat{V}_{\text{space-spin}}$ the single-particle density $n(z)$ is
$\langle\phi_{B0}|\hat{U}^{-1}\hat{n}(z)\hat{U}|\phi_{B0}\rangle$ where 
$\hat{n}(z)=\sum_{j}\delta(z-z_{j})$. The state $|\phi_{B0}\rangle$ is
a simultaneous eigenstate of the longitudinal spin operator
$\hat{S}_{z}=\sum_{j}\hat{s}_{jz}$, which has eigenvalues
$S_{z}=-\frac{1}{2}N,-\frac{1}{2}N+1,\cdots,\frac{1}{2}N-1,\frac{1}{2}N$.
The ground state of $\hat{H}_{1D}^{B}$, which now includes
$\hat{V}_{\text{space-spin}}$, is $\hat{U}|\phi_{B0}\rangle$,
and the $(N+1)$-fold degeneracy is not lifted by the magnetic field
gradient so long as the magnetic field vanishes at $z=0$. One has  
$\hat{U}^{-1}\delta(z-z_{j})\hat{U}
=\delta(z-z_{j}-\frac{\mu_{B}c}{m\omega_{\text{long}}^{2}}\hat{s}_{jz})$,
whose expectation value is $N^{-1}n_{0}(z\mp z_{0})$ when the eigenvalue of
$\hat{s}_{jz}$ is $\pm\frac{1}{2}$, where 
$z_{0}=\frac{\mu_{B}c}{2m\omega_{\text{long}}^{2}}$. If the eigenvalue of
$\hat{S}_{z}$ is $S_z$ then $wN$ of the $\hat{s}_{jz}$
have eigenvalue $\frac{1}{2}$ and $(1-w)N$ have eigenvalue $-\frac{1}{2}$,
where the fraction $w$, which satisfies $0\le w\le 1$, is
$w=N^{-1}S_{z}+\frac{1}{2}$. It follows that the single-particle density
$n(z)$ in a ground state $\hat{U}|\phi_{B0}\rangle$ with longitudinal spin 
$S_z$ is a weighted average of the extremal densities: 
$n(z)=wn_{0}(z-z_{0})+(1-w)n_{0}(z+z_{0})$. The ground state wave function
of the corresponding spinor Fermi gas differs by a factor $A$, 
the previously-given mapping function. It has the same longitudinal
spin eigenvalues and same degeneracy, and since $A^{2}=1$ these fermionic
ground states have the same density profiles as the bosonic ones.
\begin{acknowledgments}
We are very grateful to Doerte Blume for helpful comments and for 
communications regarding her closely-related
works with Brian Granger \cite{GraBlu03} and with Astrakharchik {\it et al.} 
\cite{AstBluGioPit03}.
This work was supported by Office of Naval Research grant N00014-03-1-0427
(M.D.G. and M.O.) and by NSF grant PHY-0301052 (M.O.).
\end{acknowledgments}
\end{document}